# On the Cut-off Distance and the Classical Energy-averaged Electron-ion Momentum Transport Cross Section in Ideal and Nonideal Plasmas

Mofreh R. Zaghloul

*Abstract*—The wide use of a *speed-independent* distance as a cut-off impact parameter together with Rutherford's scattering formula, within the cut-off theory, to account for charge screening in plasma environment embodies a clear inconsistency. A new physically justified choice of the cut-off distance is introduced and used to derive a closed form expression for the effective momentum transport cross section. A simple approximation for the present Coulomb logarithm, free of special functions, is also presented and assessed. A comparison with experimentally recovered data for the reduced Coulomb conductivity showed better agreement and better physical behavior of the present expressions compared to previous cut-off expressions in the literature.

## I. INTRODUCTION AND BACKGROUND

A plasma behaves ideally (like a mixture of ideal gases) when the interaction potential energy between charged particles is negligible compared to the mean thermal energy of the system; hence an ideal plasma satisfies the criterion of ideality;

$$\frac{e^2 (n_e + n_i)^{1/3}}{4 \pi \varepsilon_0 K_B T} << 1. \tag{1}$$

where $n_e$ is the number density of free electrons, $n_i$ is the number density of ions of all multiplicities, $K_B$ is the Boltzmann constant, and $T$ is the absolute equilibrium temperature. On the other hand, when the above criterion of ideality is not satisfied, correlations among plasma particles become important and the plasma deviates from the ideal behavior and is classified to be nonideal.

M. R. Zaghloul is with the Department of Physics, College of Sciences, United Arab Emirates University, Al-Ain, POB 15551, UAE. (corresponding author phone: 971-3-;713-6324  e-mail: M.Zaghloul@uaeu.ac.ae).

One of the most important problems in the calculation of transport properties of ideal and nonideal plasmas is to take account of the mechanism of charged particles interactions. Largely, this is done in terms of an effective scattering collision cross-section in the two-particle approximation. In describing the electron-ion interaction, for example, it is most common to describe the mutual interaction as an interaction between point charges under the influence of a central Coulomb force. However, for practical reasons, the infinite-range Coulomb force is cut-off at a certain appropriate distance, $r_{cut}$, such that for an electron-ion pair

$$\vec{F}(\vec{r}) = -\frac{ze^2}{4\pi\varepsilon_0 r^2}\hat{r}, \quad r \leq r_{cut}$$
$$= 0 \quad r > r_{cut} \tag{2},$$

where $ze$ is the ion charge, $r$ is the separation between the two particles and $\varepsilon_0$ is the permittivity of vacuum.

The screening of ionic potential due to presence of other, neighboring, charged particles in a plasma environment is usually accounted for through the use of a maximum or cut-off impact parameter. In the derivation of the widely used Spitzer's formula [1] the cut-off impact parameter is chosen to be the Debye-Hückel screening radius, $\lambda_D$.

The scattering problem is classically approached, as in the derivation of Rutherford's scattering formula, by considering a system of electron-ion pair that has a reduced mass $m$, electric charges $-e$, $+ze$, separated by *large* distance and which approach each other, with relative kinetic energy $E_1 = \frac{1}{2} mv^2$, at an impact parameter $b$. The situation is simply depicted as in figure 1. At position 2, the two particles are at the distance of closest approach, $r_0$, with relative kinetic energy $E_2$ and potential energy $V_2$.

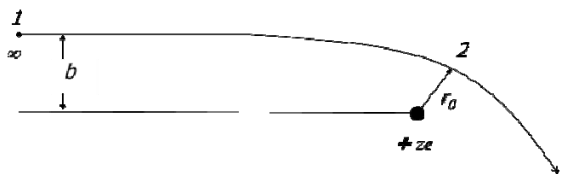

Figure 1. Descriptive trajectory of an electron undergoing a Coulomb interaction with an ion.

Applying the interaction invariants equations (conservation of energy and conservation of angular momentum) leads to the physically accepted solution or relation between the impact parameter $b$ and the distance of closes approach $r_0$

$$b = r_0 \left(1 - \frac{V_2}{E_1}\right)^{1/2}$$

$$= r_0 \left(1 + \frac{z e^2}{4 \pi \varepsilon_0 r_0 (m v^2 / 2)}\right)^{1/2} \tag{3},$$

while the scattering angle and the impact parameter maintain the relation

$$b = \frac{z e^2}{8 \pi \varepsilon_0 (m v^2 / 2)} \cot(\theta/2) \tag{4}.$$

The total momentum-transport cross-section for an azimuthally symmetric e-i collision can be recovered from the differential cross section for scattering $\sigma_{diff}(v,\theta)$ by the integration

$$Q_{ei}(v) = 2\pi \int_{\theta_{min}}^{\pi} (1 - \cos\theta)\, \sigma_{diff}(v,\theta) \sin\theta\, d\theta$$

$$= 2\pi \int_0^{b_{max}} (1 - \cos\theta)\, b\, db \tag{5},$$

where $v$ is the relative speed of approach of the two interacting particles, $\theta$ is the angle of deflection, $b$ is the impact parameter, $b_{max}$ is an upper limit of the impact parameter (cut-off impact distance) corresponding to a minimum angle of deflection, $\theta_{min}$, needed to prevent divergence of the integration in equation (5).

Considering Rutherford's scattering formula for the differential cross section for the right hand side of equation (5), and the above relation between the impact parameter and the distance of closest approach, one can write the total scattering cross section as

$$Q_{ei}(v) = 16\pi \left(\frac{z e^2}{8 \pi \varepsilon_0 m v^2}\right)^2 \ell n \left(\frac{\sin(\theta_{max}/2)}{\sin(\theta_{min}/2)}\right) \tag{6},$$

where $\theta_{max}$ and $\theta_{min}$ are the maximum and minimum deflection angles, which correspond to the minimum and maximum impact parameters, respectively.

The well-known singularity of the total cross section of Coulomb scattering by a point charge arises from scattering through small angles ($\sin(\theta_{min}/2) \rightarrow 0$). Hence, a minimum scattering angle, $\theta_{min}$, needed to be specified to avoid the divergence of the cross section, which is equivalent to setting the differential scattering cross section equal to zero for angles smaller than $\theta_{min}$.

## II. Cut-off Distance

Using the relations (3) and (4), it is easy to show that

$$\frac{1}{\sin(\theta/2)} = 1 + \frac{8\pi\varepsilon_0 (mv^2/2)}{ze^2} r_0 \tag{7}$$

Substituting for the scattering angle from (4) or from (7) into Eq. (6), setting $\theta_{max}=\pi$ the total cross section can be written, in order, in the following two equivalent forms

$$Q_{ei}(v) = 16\pi \left(\frac{ze^2}{8\pi\varepsilon_0 mv^2}\right)^2$$

$$\times \ln\left[1 + \left(\frac{4\pi\varepsilon_0 mv^2}{ze^2} b_{cut}\right)^2\right]^{\frac{1}{2}} \tag{8}$$

and

$$Q_{ei}(v) = 16\pi \left(\frac{ze^2}{8\pi\varepsilon_0 mv^2}\right)^2$$

$$\times \ln\left(1 + \left(\frac{4\pi\varepsilon_0 mv^2}{ze^2} r_{0,max}\right)\right) \tag{9}$$

It is important to note that in the derivation of Rutherford's scattering formula and hence in the derivation of the expressions (8) and (9), a Coulomb potential of *infinite* range has been used comprising an inconsistency that will be considered below. Within the cut-off theory, one may consider

Eq. (8) or Eq. (9) as an exact and general result that depends on one fundamental parameter; the cut-off distance, $b_{cut}$ or equivalently $r_{0,max}$, which needs to be determined for practical computation of the cross section.

The above two equations give the classical momentum transport cross section for elastic scattering although usually written in literature in the form (8) (see for example Mitchener and Kroger [2]). The two forms (8) and (9) are fundamentally the same and are essentially equal for equivalent cut-off distances. A first look at Eq. (8) suggests the use of a fixed cut-off impact parameter as widely used in the literature (though not necessarily) while a first look at Eq. (9) suggests the use of a constant cut-off distance of closest approach (though not necessarily too).

According to Eq. (4) the use of a fixed cut-off impact parameter could in principle remove important scattering interactions with large angles while counting scatterings with small angles, in the same time, depending on the relative speed of approach $v$. This is admittedly a concern in a plasma environment where free electrons may assume any velocity and are generally described by a velocity distribution function. Such a choice of the cut-off distance is adopted in Spitzer's model for ideal plasma environment and in many similar models in the literature. Consequently, removing small angle deflections which cause the singularity of the cross section would require a cut-off impact parameter for each speed, which is not the case when a fixed, *speed-independent*, impact parameter is chosen as the cut-off distance as above. Better understanding of the appropriate choice of the cut-off distance can be reached by manipulating Eqs. (4) and (7) to get

$$r_0 = \left(\frac{1-\sin(\theta/2)}{\cos(\theta/2)}\right) b = f(\theta/2) b \tag{10}.$$

This relation is valid for all approaching speeds, so it is independent of $v$. However, it clearly shows that for small scattering angles the distance of closest approach $r_0$ approaches the impact parameter $b$. Since the factor $f(\theta/2)$ is bound by a maximum value of unity, then cutting-off at a large fixed value of $r_0$ (i.e. cutting-off all values of $r_0 > r_{0,max}$) will certainly imply cutting-off large values of $b$ with small values of $\theta$. However, and again cutting-off at a constant value of $b$ does *not* guarantee cutting-off small angles and even worse it could cut large angle encounters as large angles and small values of $r_0$ are possible causes for the large value of $b$.

Accordingly, in the present model we recommend cutting-off scatterings leading to distances of closest approach greater than $r_s$ where $r_s$ is a characteristic screening distance. It has to be noted that this choice of the cut-off distance corrects for the inconsistency buried in the above treatment of the scattering problem where a *finite-range* Coulomb force is sought while an *infinite-range* Coulomb force is considered through the whole projectile trajectory from very large distance (infinity) to the neighborhood of the scattering center. To explain this, we consider Fig. 2 which shows two types of projectile trajectories; 1- using the infinite Coulomb force like the two trajectories represented by the thick dashed curves in the figure with impact parameters $b_0(b_1)$ and $b_0(b_{1,cut})$, and 2- using a finite-range Coulomb force as given in Eq. (2) like the trajectories represented by the two thin solid curves with impact parameters $b_1$ and $b_{1,cut}$.

A projectile of impact parameter $b_0$ under the influence of infinite-range Coulomb force hits a sphere of radius $r_{cut}$ (interaction volume for the *finite-range* Coulomb force) at the position labeled 1 with a velocity tangent to the trajectory and speed $v_1$. Within the interaction volume, this trajectory is in every respect identical to the trajectory of impact parameter $b_1$, and speed of approach $v_1$ using the finite-range Coulomb force where the projectile moves in a force-free region with a constant speed in a straight line till entering the interaction volume where it is affected by the Coulomb force.

The diagram simply shows that for the *finite-range* Coulomb force, trajectories with impact parameters greater than the *fixed value* $b_{1,cut}=r_{cut}$ will not interact with the scattering center (correspond to a zero-scattering angle) and hence the choice of a fixed cut-off impact parameter is acceptable in this case (case of *finite-range* Coulomb force). However, this is not the case with the infinite Coulomb force. Mapping such a distance to the corresponding trajectory from the *infinite-range* Coulomb force shows that the corresponding cut-off impact parameter is greater than the cut-off distance of the force, namely $r_{cut}$, and is given by $b_0(r_{cut})$ or $b_0(b_{1,cut})$. Applying the interaction invariants, it is trivial to show that

$$b_0 = b_1 \left( 1 + \frac{z\,e^2}{4\,\pi\,\varepsilon_0\,r_{cut}\,(mv^2/2)} \right)^{1/2}$$

(11).

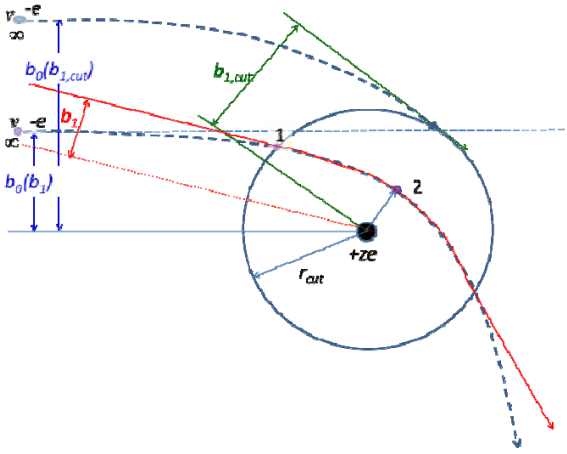

Figure 2. Descriptive trajectories of an electron undergoing a Coulomb interaction with an ion using an infinite-range Coulomb force (thick dashed curves) and a cut-off Coulomb force (thin solid curves).

Comparing this expression with Eq. (3) provides with sufficient clarity the justification for our choice of the cut-off distance as $b(r_{cut})$ or $b(r_s)$ where $r_s$ is a fixed distance independent of the speed of approach which represents the cut-off distance or cut-off impact parameter for a finite-range Coulomb force as given by Eq. (2). This choice of the cut-off distance within the present analysis resolves the inconsistency of using an infinite-range force in obtaining trajectories while seeking to cut-off the force in the same time.

III. **Energy-Averaging**

Accurate formulae for a multicomponent gas mixture under thermal and chemical equilibrium may be obtained from the Chapman-Enskog approximation. When the deviation from the Maxwell–Boltzmann distribution in the equilibrium is small, the energy-averaged momentum transport cross section can be expressed as [3,4,5]

$$\overline{Q}_{ei} = \frac{8}{3\sqrt{\pi}} \left(\frac{m}{2 K_B T}\right)^{5/2} \left(\frac{\pi m}{8 K_B T}\right)^{1/2}$$
$$\times \int_0^\infty v^5 Q_{e-i}(v) \exp\left(\frac{-m v^2}{2 K_B T}\right) dv$$

(12).

Substituting from Eq. (9) into Eq. (12) and using the substitutions, $x = m v^2 / 2 K_B T$, $\bar{b}_0 = z e^2 / 12 \pi \varepsilon_0 K_B T$, $\Lambda_s = r_s / \bar{b}_0$, the above integration can be evaluated analytically giving rise to

$$\bar{Q}_{e-i} = 6 \pi \bar{b}_0 \int_0^\infty e^{-x} \ln\left[1 + \Lambda_s \left(\frac{2x}{3}\right)\right] dx$$
$$= 6 \pi \bar{b}_0 \exp(\eta) E_1(\eta) \quad (13),$$

where $\eta = 3/(2\Lambda_s)$, $E_1$ is the exponential integral and $\bar{b}_0$ represents physically the impact parameter for a $90^0$ scattering of a particle having the average thermal energy ($3/2\ K_BT$).

Thus, the resulting expression from the present model to replace the ordinary Coulomb logarithm, in the computations of transport properties of ideal and nonideal plasma, becomes

$$\ell n \Lambda \rightarrow \frac{\bar{Q}_{ei}}{6 \pi \bar{b}_0^2} = \exp\left(\frac{3/2}{\Lambda_s}\right) E_1\left(\frac{3/2}{\Lambda_s}\right) \quad (14).$$

It is worthy to mention that the exact expression derived previously from the classical cut-off theory for interaction of point charges [6], namely,

$$\ell n \Lambda \rightarrow \frac{\bar{Q}_{ei}}{6 \pi \bar{b}_0^2} = \frac{\pi}{2} \sin(\beta)$$
$$\times \left[1 - \frac{2}{\pi}\left(Si(\beta) + Ci(\beta)/\tan(\beta)\right)\right], \quad (15),$$

where $\beta = 3\bar{b}_0 / 2 r_s$, can be easily derived by energy averaging Eq. (8) setting $b_{cut}=r_s$ (i.e. fixed impact parameter) or by energy averaging Eq. (9) setting $r_{0,max}$ to

$$r_{0,max} = \frac{z e^2}{4 \pi \varepsilon_0 m v^2}\left(-1 + \sqrt{1 + \frac{4 \pi \varepsilon_0 m v^2}{z e^2} r_s^2}\right) \quad (16).$$

However, as explained above, such a choice of the cut-off distance is not consistent with the assumed finite range Coulomb force.

## IV. Quantitative assessment

For the purpose of comparison and assessment one may simply follow Spitzer's choice of $\lambda_D$ as the screening distance in a plasma system with $\Lambda_s = \Lambda_D = \lambda_D/\bar{b}_0$ and compare the Coulomb logarithm as calculated from three different models; (a) from Spitzer's classical expression, (b) from the present model (cut-off distance of closest approach Eq. (14)), and (c) from the cut-off impact parameter model Eq. (15).

Figure 3 shows this comparison in terms of the reduced Coulomb conductivity, $\sigma_c^*$ (inverse of the Coulomb logarithm) as a function of the Debye nonideality parameter $\Gamma_D = 3/\Lambda_D$. The results constitute a set of universal curves for the reduced Coulomb conductivity. Although Spitzer's expression works well for ideal plasmas, it diverges in the nonideal regime ($\Gamma_D > 1.0$). However, the reduced conductivity deduced from the present model (Eq. (14)) shows less divergence, at high degrees of nonideality, and hence better performance and better agreement with values recovered from experimental measurements compared to those deduced from equation (15) and from Spitzer's formula.

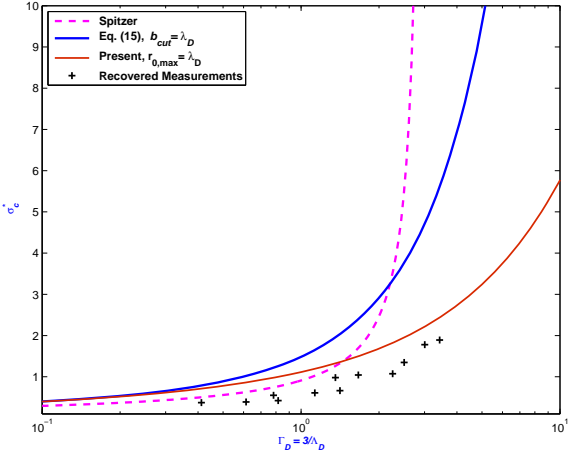

Figure 3. Reduced Coulomb conductivity, $\sigma_c^*$ (inverse of Coulomb logarithm) as a function of the Debye nonideality parameter $\Gamma_D = 3/\Lambda_D$, as calculated from exact formulae for point charges interactions and from the finite size formula using $\Lambda_z=0.01$ as functions of the Debye nonideality parameter $\Gamma_D = 3/\Lambda_D$; Measurements by Ivanov et al [5] taken from [6].

## V. A Simple Approximation to the Coulomb Logarithm

As has been shown above, the exact formula introduced herein for the Coulomb logarithm (Eq. (14)) has been expressed in terms of the exponential integral. Although many software packages have algorithms to calculate values of special functions such as the exponential integral, it is always desirable or sometimes necessary to calculate plasma transport properties with a minimum of computational effort. In a previous work [7], it was shown that the quantum Coulomb logarithm, which has been also expressed in terms of the exponential integral, can be approximated by the simple Spitzer logarithmic expression with the aid of a parameter, $\xi = 0.908956701$. This enables one to derive the following approximation for the present Coulomb logarithm where

$$\ell n \Lambda \approx \frac{1 + \ell n \left(1 + 0.1377\, \Lambda_s\right)}{1.0 + 1.5 / \Lambda_s} \tag{17}$$

Figure 4 shows the percentage of the absolute of the relative error resulting from using the approximate Coulomb logarithm given by equation (17). As it can be seen from the figure, the approximate expression (17) can be satisfactorily used with a maximum error less than 5%. In addition, the above expression simply reduces to $\ell n\, \Lambda_s$ as $\Lambda_s \to \infty$.

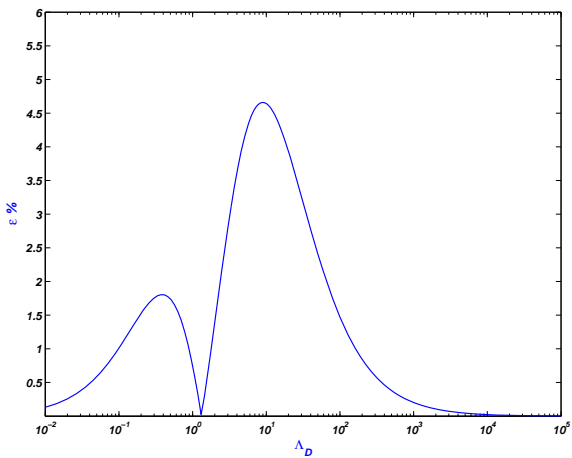

Figure 4 Absolute of percentage relative error of the approximate formula for the Coulomb logarithm, Eq. (17).

## VI. Conclusion

The choice of the cut-off impact parameter has been revisited and examined showing inconsistency and inaccuracy of the commonly used fixed *speed-independent* value. A new physically justified choice of the cut-off distance has been introduced and used to derive an exact analytical expression for the energy-averaged momentum transport cross within the cut-off theory. The derived expression is valid for both ideal and nonideal plasma. A simple approximation for the newly derived Coulomb logarithm, free of special functions, has been also introduced and assessed.

Acknowledgements

The author wishes to thank the anonymous reviewers for useful discussion and helpful suggestions.

1818